\documentstyle[multicol,prl,aps]{revtex}
\begin{document}

\draft

\title{Mesoscopic Kondo Effect in an Aharonov-Bohm Ring}

\author{Kicheon Kang$^1$ and Sung-Chul Shin$^2$}
\address{ $^1$Department of Physics, Chonbuk National University,
            Chonju 561-756, Chonbuk, Korea}
\address{ $^2$Department of Physics and Center for Nanospinics of Spintronic
          Materials, Korea Advanced Institute of Science and Technology,
          Taejon 305-701, Korea }  

\date{\today}

\maketitle

\begin{abstract}
An interacting quantum dot inserted in a mesoscopic ring is
investigated. A variational ansatz is employed to describe the ground
state of the system in the presence of the Aharonov-Bohm flux. 
It is shown that, for even number of electrons
with the energy
level spacing smaller than the Kondo temperature, the persistent current
has a value similar to that of a perfect ring with the same radius. 
On the other hand, for a ring with odd number electrons, the persistent current
is found to be 
strongly suppressed compared to that of an ideal ring, which implies
the suppression of the Kondo-resonant transmission.
Various aspects of the Kondo-assisted persistent current are discussed.
\end{abstract}
\pacs{PACS numbers: 
           72.15.Qm, 
           73.23.Ra, 
           73.23.Hk  
 }
%
\begin{multicols}{2}
The Kondo effect is one of the most exciting and intensively studied
problems in condensed matter physics~\cite{hewson93}. Recent remarkable progress
in the nano-fabrication of electronic devices made it possible 
to investigate the Kondo effect by means of a quantum dot (QD) coupled
to electrodes by tunneling barriers~\cite{gordon98,cronen98,simmel99}.  
In contrast to the conventional Kondo problem of magnetic impurities in
a metal host, the QD device enables more detailed and systematic study
on the Kondo effect.  Number of electrons and level position of the QD, 
and coupling strength of the leads to the QD are made controllable, so
that a ``tunable" Kondo effect could be realized.  
The hallmark of the Kondo effect in tunneling devices 
is the enhanced conductance due to the Kondo 
resonance~\cite{glazman88,ng88,hershfeld,meir,yeyati93,konig96,kang98,craco99}. 
The Kondo resonance at the Fermi level provides a new channel for electric
current flowing through the QD.
On the other hand, it will be also interesting to study on what happens to
the Kondo effect where the size of the metallic host itself
is mesoscopic. For a single Kondo impurity embedded in an ultra-small
metallic grain, the Kondo resonance is strongly affected when the
mean level spacing ($\delta$) 
in the grain becomes larger than the bulk Kondo 
temperature ($T_K^0$), in a way that depends on the parity of the number
of electrons on the grain~\cite{thimm99}.

In this Letter, we study theoretically a quantum dot embedded in a mesoscopic 
Aharonov-Bohm (AB) ring. This geometry has been the subject of 
Ref.\cite{buttiker96,ferrari99}. Actually this is another 
example of a Kondo impurity embedded in a mesoscopic host.
We discuss the Kondo effect present in 
the behavior of the persistent current,
in relation to: (i) the level discreteness of the ring,
(ii) the phase coherence of the Kondo resonance, and (iii) the parity, i.e.,
the number of electrons of the system (QD + ring) being even or odd. 
A previous theoretical 
study on the Kondo impurity in a mesoscopic ring has been 
reported~\cite{ferrari99}, which does not seem to provide proper 
descriptions on the effect of the 
level discreteness and the parity.
We find that, for even parity with small level spacing $\delta\lesssim T_K^0$, 
the magnitude of the persistent
current $I$ is of the same order of $I_0$$(=ev_F/(2\pi R)$, where $v_F$
and $R$ are the Fermi velocity and the radius of the ring, respectively.), 
the value found in a perfect ring with the same size.
This implies a high transmission through
the QD due to the Kondo resonance. We show that transmission through the QD is
reduced in the $\delta > T_K^0$ limit, so is $I/I_0$.
Surprisingly, the odd-parity ring shows quite different behavior.
The persistent current is completely suppressed compared to $I_0$ in the
small $\delta$ limit. It shows a relationship $I/I_0 \sim \delta/T_K^0$
at $\delta\ll T_K^0$, which means a very low 
transmission through the quantum dot in the Kondo limit. 
It is also important to notice that the ground state
depends on the AB flux in the mesoscopic region.
In other words, the Kondo effect depends on the phase of hopping amplitude
between
the QD and the ring while its magnitude remains unchanged. In this sense, 
the Kondo effect in a small ring is phase-coherent,
and can be called a {\em mesoscopic Kondo effect}.
The phase-dependence of the Kondo effect disappears at thermodynamic limit,
as one may expect.
 
A QD embedded in an AB ring is described by a one-dimensional
tight-binding Hamiltonian with the nearest neighbor hopping amplitude $t$
of $N$ lattice sites including an Anderson impurity:
$ H = H_0 + H_D + H_T $,  
where $H_0$, $H_D$, and $H_T$ represent the noninteracting part of the
ring, the QD (or an Anderson impurity), and tunneling,
respectively. This Hamiltonian can be written in the following simplified
form by using the diagonalized basis of $H_0$:
\begin{eqnarray}
 H &=& \sum_{m\sigma} \varepsilon_m c_{m\sigma}^\dagger c_{m\sigma} 
       + \sum_\sigma \varepsilon_d d_\sigma^\dagger d_\sigma
       + U \hat{n}_\uparrow \hat{n}_\downarrow \\  
  &+&  \sum_{m\sigma} \left( t_m c_{m\sigma}^\dagger d_\sigma 
             + \mbox{\rm h.c.} \right) \nonumber \; .
\end{eqnarray}
$\varepsilon_d$ and $U$ represent the single particle energy and the
Coulomb repulsion in the QD, respectively. 
$ \varepsilon_m  = -2t \cos{\frac{m}{N}\pi} $, 
$ t_m = \sqrt{\frac{2}{N}}\,i\,\sin{\frac{m}{N}\pi}
     \left( t_L + t_R e^{i\phi} (-1)^{m+1} \right) $ 
with $m=1,2,\cdots N-1$.
Here $t_L$ ($t_R$) denotes the hopping matrix element of the quantum dot
state with the left (right) neighboring site. The phase factor
$\phi$ is defined by $\phi=2\pi \Phi/\Phi_0$, where $\Phi$ and
$\Phi_0$ are the external flux and the flux quantum ($=hc/e$),
respectively.
An important parameter describing the coupling strength between
the QD and the ring, namely $\Gamma$, is defined by 
\begin{displaymath}
 \Gamma = \pi \rho(\varepsilon_F) |t(\varepsilon_F)|^2 ,
\end{displaymath}
where $\rho(\varepsilon_F)$ and $t(\varepsilon_F)$
represent the density of states and the hopping amplitude
at the Fermi energy $\varepsilon_F$, respectively.
For half-filled case ($\varepsilon_F=0$) at continuum limit,
$\rho(0)=\delta^{-1}=N/(2\pi t)$ and $|t(0)|^2=\frac{2}{N} (t_L^2+t_R^2)$.
Thus, $\Gamma$ is given by 
\begin{equation}
 \Gamma = \frac{t_L^2+t_R^2}{t} .
\end{equation}

One of the simplest ways of investigating this Hamiltonian in a 
controlled manner is
to adopt a variational ansatz~\cite{gunnar87}.
Here we consider only infinite-$U$ case because including the effect of
finite-$U$ doesn't provide any essential change in the Kondo limit
except for the value of the Kondo temperature~\cite{kang96}.
For infinite-$U$, the $N_e$-particle variational ground state 
$|\Psi(N_e)\rangle$ 
is given as follows in the leading order of
$1/N_s$-expansion with 
$N_s$ being the magnetic degeneracy (equals to two in our case),
\begin{equation}
 |\Psi(N_e)\rangle = A \left( |\Omega(N_e)\rangle 
   + \frac{1}{\sqrt{2}} \sum_{m\sigma}^{occ}
     \alpha_m |\varepsilon_m\sigma(N_e)\rangle \right) 
   \;,
\end{equation}
where
$|\Omega(N_e)\rangle$ denotes $N_e$-particle ground state of $H_0$,
and $|\varepsilon_m\sigma(N_e)\rangle = d_\sigma^\dagger c_{m\sigma}
|\Omega(N_e)\rangle$. The summation in the second term of the R.H.S. is to be
taken for the occupied levels of $H_0$. 
It is known that this 
variational ground state describes well the essence of the Kondo
physics~\cite{gunnar87}. 
The constants $A$ and $\alpha_m$ are 
determined by minimization of the function 
$\langle\Psi(N_e) |H| \Psi(N_e)\rangle - 
E_0 ( \langle\Psi(N_e)|\Psi(N_e)\rangle - 1 )$
with respect to $A$ and $\alpha_m$.
As a result, the ground state energy $E_0$ can be achieved from
the self-consistent equation
\begin{equation}
 E_0' = \sum_{m\sigma}^{occ} \frac{|t_m|^2}{ E_0'+\varepsilon_m-
   \varepsilon_d} \; , \label{eq:energy}
\end{equation}
where $E_0'\equiv E_0-E_\Omega$ with $E_\Omega$ being the energy
of the noninteracting ground state $|\Omega(N_e)\rangle$.
The ``correlation energy" $T_K$, which characterizes the ground state,
is defined by the 
energy lowering due to coupling between the impurity and the host:
\begin{equation}
 T_K = \varepsilon_d - \varepsilon_F - E_0' \;,
 \label{eq:kondo} 
\end{equation}
where
$\varepsilon_F$ is the energy of the highest occupied level which corresponds
to the Fermi energy at thermodynamic limit.
$T_K$ corresponds to the Kondo temperature, namely $T_K^0$, in the bulk 
($\delta\rightarrow 0$) Kondo limit.

Fig.\ref{fig:kondo} displays the correlation energy as a function 
of the level spacing $\delta$ for several parameter sets in the
Kondo limit ($\varepsilon_d < -2\Gamma$). 
We chose $\phi=\pi/2$, where the cross term proportional to $t_Lt_R$
 in $|t_m|^2$ vanishes.  
Eq.(\ref{eq:energy}) depends 
on the parity and leads to a parity-dependent behavior of the
correlation energy. $T_K$ shows a highly accurate scaling 
behavior in a parity dependent manner
with the characteristic energy $T_K^0$, the Kondo temperature.
Note that the value of $T_K^0$ itself depends
strongly on the given parameters. 

In a quantum dot coupled to two semi-infinite
electrodes, Kondo resonance leads
to an enhanced transmission through the quantum dot. The transmission
becomes perfect for a symmetric coupling in the Kondo limit, $\varepsilon_d
\ll-2\Gamma$~\cite{glazman88,ng88}.
In our case of a QD embedded in an AB ring, transmission through 
the QD is closely related to the persistent current. One may
expect that the Kondo effect gives rise to a high transmission through
the QD as in the case of the open system, and accordingly, leads to
a persistent current with its magnitude comparable to $I_0$.
However, the result is somewhat different from this simple expectation 
as we show in the following.

At zero temperature the persistent current is given by
\begin{mathletters}
\begin{equation}
 I = -\frac{e}{\hbar} \frac{\partial E_0}{\partial\phi} \; .
  \label{eq:curr}
\end{equation}
Since the variables $\varepsilon_d$, $\varepsilon_F$, and $E_\Omega$
in Eq.(\ref{eq:kondo}) 
are all independent of the phase $\phi$, Eq.(\ref{eq:curr}) can
be rewritten as
\begin{equation}
 I = \frac{e}{\hbar} \frac{\partial T_K}{\partial\phi} \; .
 \label{eq:currTK}
\end{equation}
\end{mathletters}

By using the self-consistent equation for the ground state energy
in Eq.(\ref{eq:energy}) we obtain the following expression for 
the current:
\begin{mathletters}
\begin{equation}
 I = \frac{e}{\hbar} A^2 \sum_{m\sigma}^{occ} \frac{u_m}{E_0'+\varepsilon_m
   -\varepsilon_d} \sin{\phi} \;,
 \label{eq:curr2}
\end{equation}
where the renormalization constant is given by
\begin{equation}
 A^2 = \left[ 1 + \sum_{m\sigma}^{occ} \frac{|t_m|^2}{ (E_0'+\varepsilon_m-
   \varepsilon_d)^2 } \right]^{-1} \;,
\end{equation}
and
\begin{equation}
 u_m = \frac{4}{N} (-1)^{m+1} \sin^2{\left(\frac{m}{N}\pi\right)} t_L t_R \; .
\end{equation}
\end{mathletters}

Fig.\ref{fig:Id} shows numerical result of the
current vs. energy level spacing. Highly
accurate scaling result is found as in the case of the correlation energy. 
For even parity, $I/I_0\simeq \pi/2$ is constant at $\delta\ll T_K^0$, which
implies a high transmission through the QD. Note that $I_0=ev_F/(2\pi R)
=\frac{e}{h}\delta$, and thus, $I$ is also proportional to $\delta$ 
in this region. It is interesting to note that $I/I_0=\pi/2$ together with the
sinusoidal current-phase relation implies that $E_0(\phi=\pi)-E_0(\phi=0)$
has the same value with the counterpart of an ideal ring.
$I/I_0$ begins to decrease around $\delta\sim T_K^0$ as 
$\delta$ is increased, which can be interpreted as
a suppression of the Kondo-resonant
transmission due to finite level spacing.
Note that the result for $\delta > T_K^0$ contradicts the conclusion
of Ref.~\cite{ferrari99}. The authors in \onlinecite{ferrari99} predict
a counterintuitive result that the persistent current will be 
enhanced due to the sharp Kondo peak even compared to an ideal ring, 
in the limit of $\delta\gg T_K^0$. However, in providing their argument, 
they did not
take into account the fact that the Kondo resonance is strongly suppressed
in this limit due to finite size of the ring. As shown here in Fig.2,
Kondo-assisted persistent current is suppressed in the $\delta > T_K^0$
limit that implies a suppression of the Kondo resonance (see also
Ref.~\onlinecite{thimm99}).

Strikingly, the situation is quite different in a ring with
odd number of electrons. In the low $\delta$ limit ($\delta\ll T_K^0$), 
the numerical result shows that $I/I_0 \propto
\delta/T_K^0$ (see the inset of Fig.\ref{fig:Id}). This implies that
the transmission through the QD is suppressed
at small $\delta$ limit in spite of the Kondo resonance. The origin of
this suppression and the linear scaling can be explained as follows.
The persistent current $I$ can be divided into two terms as
$I=I_\uparrow+I_\downarrow$, where $I_\uparrow$ and $I_\downarrow$ are
contributions from the spin-up electrons and from the
spin-down electrons, respectively.
It is important to note that $I_\sigma$ satisfies the Leggett's 
theorem~\cite{leggett91}: The direction of the persistent current
depends on the parity of the 
electron numbers with the given spin $\sigma$, $N_e^\sigma$.
For even number of $N_e$, $N_e^\uparrow =N_e^\downarrow$ and 
$I_\uparrow=I_\downarrow$, that is, two terms contribute
equally to the Kondo-resonant transmission. However, for odd parity,
$N_e^\uparrow = N_e^\downarrow+1$ or $N_e^\uparrow = N_e^\downarrow-1$.
Consequently two terms have opposite sign, but with almost the same 
magnitude because of the correlated nature of the ground state. 
Electrons of each spin class give the Kondo-assisted 
transmission, but they are almost canceled out. The only portion
which is not canceled is about $\delta/T_K^0$. 
This explains the linear-scaling
behavior of $I/I_0$ (quadratic scaling of $I$) for odd parity.  
It should be noted
that this kind of suppression does not occur for a noninteracting ring
with spin degeneracy~\cite{loss91}, and is a peculiar nature of 
the strongly correlated ground state.
This linear scaling breaks down at around $\delta = T_K^0$ and $I/I_0$
begins to decrease at $\delta/T_K^0\sim 2.4$. 

As one can see in Eq.(\ref{eq:currTK}), non-zero persistent current
indicates that the correlation energy is phase-dependent. 
In other words, Kondo effect in a small AB ring is really {\em mesoscopic}
since the characteristic energy scale of the system
depends on the phase of the transmission amplitude while its magnitude
is unaffected.
Fig.\ref{fig:Iphi} shows the current-phase relation for various values
of $\delta$. We show only the case of diamagnetic response
($dI/d\phi < 0$ at $\phi=0$). Actually, being diamagnetic or paramagnetic
is dependent on the symmetry of the highest occupied level of $H_0$, but its
current-phase relation is not affected by the symmetry except for the direction
of the current.  It is found that the current-phase 
relation is close to sinusoidal at $\delta/T_K^0\ll 1$. 
On the other hand, for $\delta/T_K^0\gg 1$
the current-phase relation is far from sinusoidal
and suffers a rapid drop at around $\phi=\pi$ (at $\phi=0$ for paramagnetic
case). This behavior is similar to the one
 originated from gap opening due to weak 
disorder near the degeneracy point of single particle energies
\cite{buttiker83}.

Fig.\ref{fig:Iene} displays the persistent current as a function of
the QD level for $\delta=0.001t$. Experimentally, the QD level can
be controlled by attaching a gate electrode to the QD.
The current shows an asymmetric peak structure. First, let us discuss 
the case of even number of electrons (Fig.\ref{fig:Iene}(a)).
As discussed above,
the Kondo-assisted tunneling leads to a high transmission through
the QD for even parity if $\delta\lesssim T_K$.  
At $\varepsilon_d\ll -2\Gamma$, the current is suppressed because 
$T_K\ll\delta$. In this limit, the correlation energy is too small to 
show a Kondo-assisted tunneling. By increasing $\varepsilon_d$, the
correlation energy is increased exponentially and the current reaches
its maximum at a point that satisfies the condition $T_K>\delta$ 
and $n_d\simeq1$. Further increasing $\varepsilon_d$ results in
a decrease of the current because of charge fluctuation. Kondo physics
is no longer valid in this region and the transmission through the
QD begins to decrease as $n_d$ decreases.

For the case of odd parity (Fig.\ref{fig:Iene}b) the situation is quite
different. The persistent current is suppressed at $\varepsilon_d\ll 
-2\Gamma$ for the same reason as in a ring with even parity. As increasing 
$\varepsilon_d$ the magnitude of the
current increases because $T_K/\delta$ increases. However,
in contrast to the case of even parity, $I$ has its maximum value
at a lower value of $\varepsilon_d$ and begins to decrease before
going into the charge fluctuation region. It decreases so fast in this region.
This behavior is closely related to the one shown
in Fig.\ref{fig:Id}. That is, the Kondo-assisted persistent current
is strongly suppressed in an odd-parity ring with $\delta\ll T_K$.

In conclusion, we have investigated the Kondo-assisted persistent 
current flowing through a quantum dot embedded in a mesoscopic
Aharonov-Bohm ring.
It is shown that the properties of the current
are determined by a single parameter $\delta/T_K^0$ in a parity-dependent
manner.  
For a ring with even number of electrons, the Kondo effect leads to
a high transmission and the magnitude of the 
persistent current is comparable to that of an ideal ring with the same
size. 
In contrast, the persistent current is strongly suppressed in a ring
with odd number of electrons. We have discussed various aspects of the
Kondo-assisted persistent current.

We thank S. Y. Cho and G. Cuniberti for helpful discussions and comments.
This work was supported by grant No. 1999-2-11400-005-5 from the 
Basic Research Program of the KOSEF.


%
%
\begin{figure}[t]
 \caption{ Parity-dependent scaling behavior of the correlation energy
  at $\phi=\pi/2$ 
  as a function of the level spacing for $\varepsilon_d=-0.75, t_L=t_R
  =0.3 (2\Gamma=0.36)$ (solid lines), for $\varepsilon_d=-0.8, t_L=t_R
  =0.3 (2\Gamma=0.36)$ (dotted lines), and for $\varepsilon_d=-0.6, t_L=t_R
  =0.25 (2\Gamma=0.25)$ (dashed lines), respectively. The calculated
  Kondo temperature $T_K^0$ for each parameter sets are given as 
  $2.08\times10^{-3}$, $1.35\times10^{-3}$, and $7.76\times10^{-4}$,
  respectively. Note that all the energy parameters are given in unit of
  $t$. 
	   }
 \label{fig:kondo}
\end{figure}
\begin{figure}
 \caption{ Parity-dependent scaling behavior of the persistent current
  as a function of the level spacing for the same
  parameters as those used in Fig.\ref{fig:kondo}.
  The inset shows the linear scaling for the odd parity at small
  $\delta/T_K^0$. 
	   }
 \label{fig:Id}
\end{figure}
\begin{figure}
 \caption{ Current-phase relation (a) for even number of
  electrons and (b) for odd number of electrons. Three different cases
  of $\delta/T_K^0=0.25,1,4$ are shown. Other parameters are given as
  $\varepsilon_d=-0.75$, $t_L=t_R=0.3$ $(2\Gamma=0.36)$, in unit of $t$. 
     }
 \label{fig:Iphi}
\end{figure}
\begin{figure}
 \caption{ Persistent current $I$ (solid lines), correlation energy $T_K$
  (dashed lines),
  and the occupation number of the quantum dot $n_d$ (dotted lines) 
  (a) for even parity and (b) for odd parity. $\phi=\pi/2$, $\delta=0.001t$,
  and $t_L=t_R=0.3t$ ($2\Gamma=0.36$). 
   }
 \label{fig:Iene}
\end{figure}
\end{multicols}
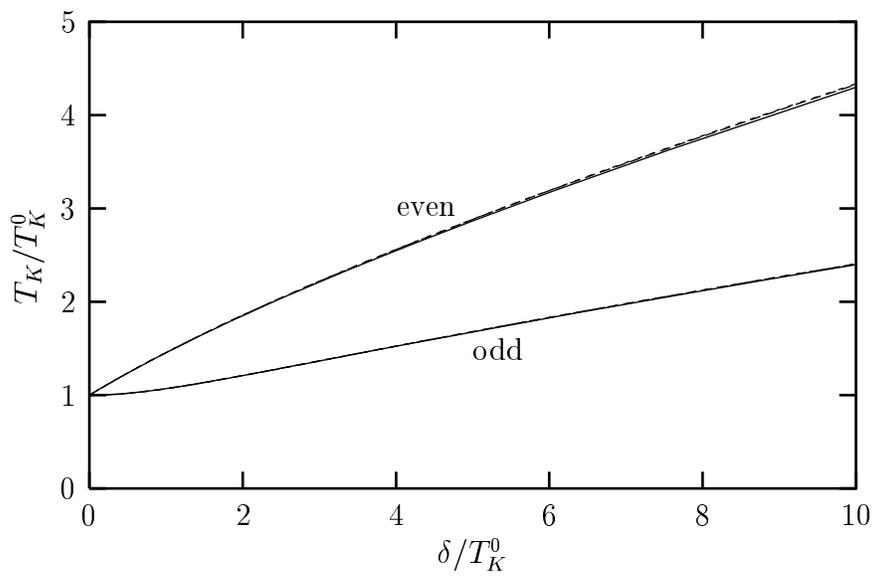
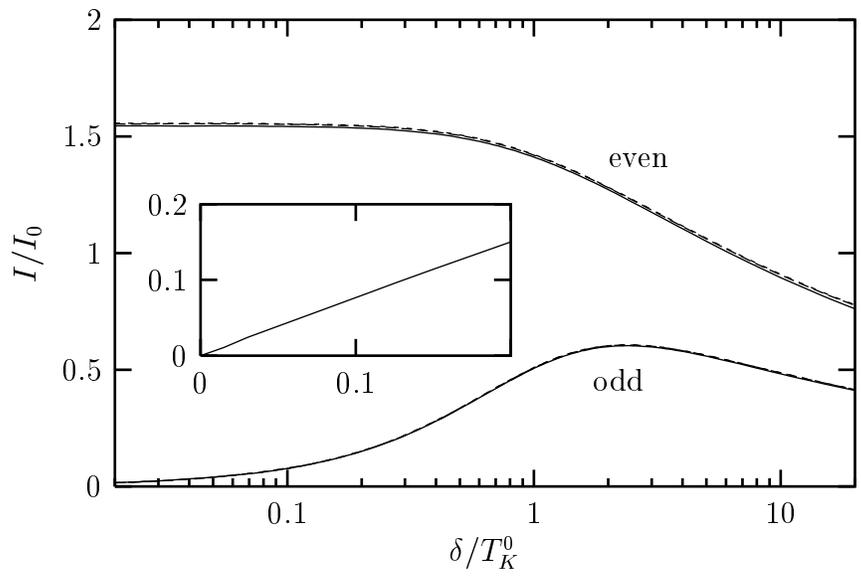
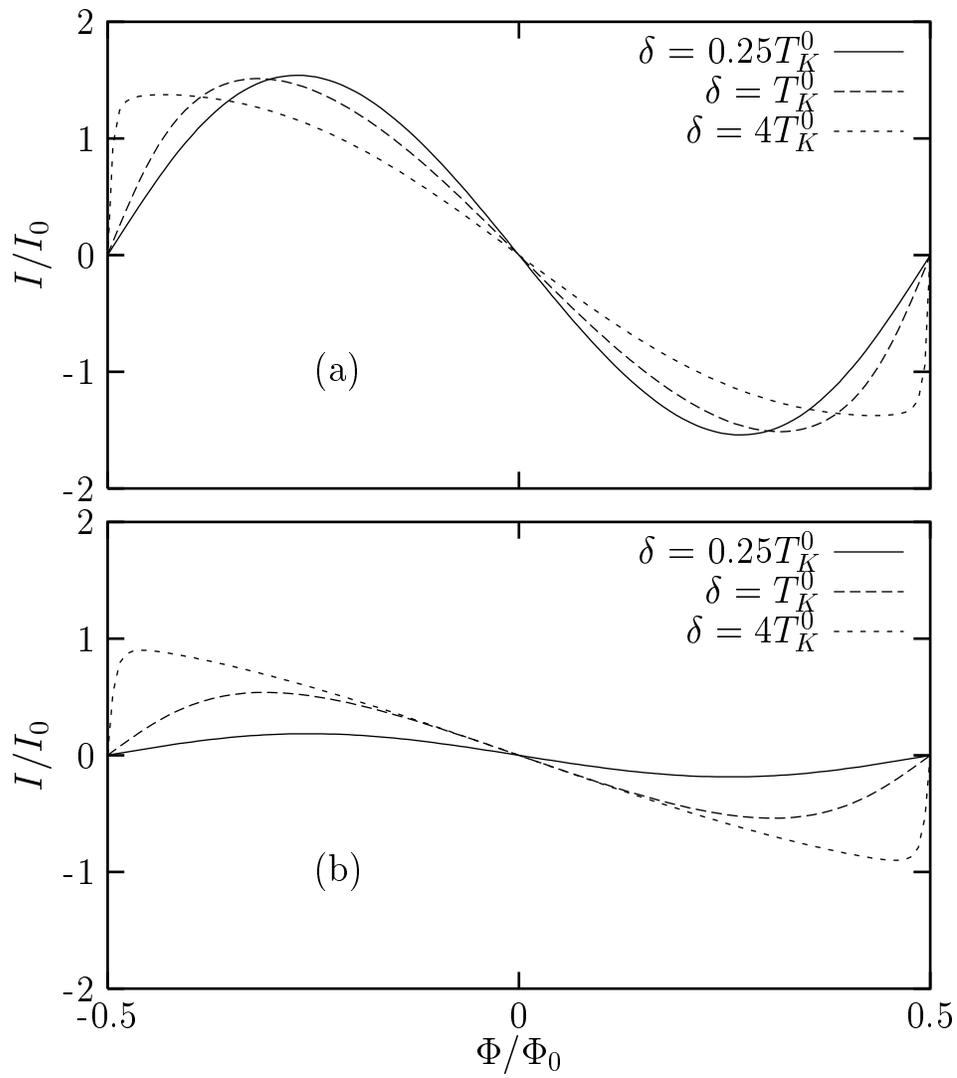
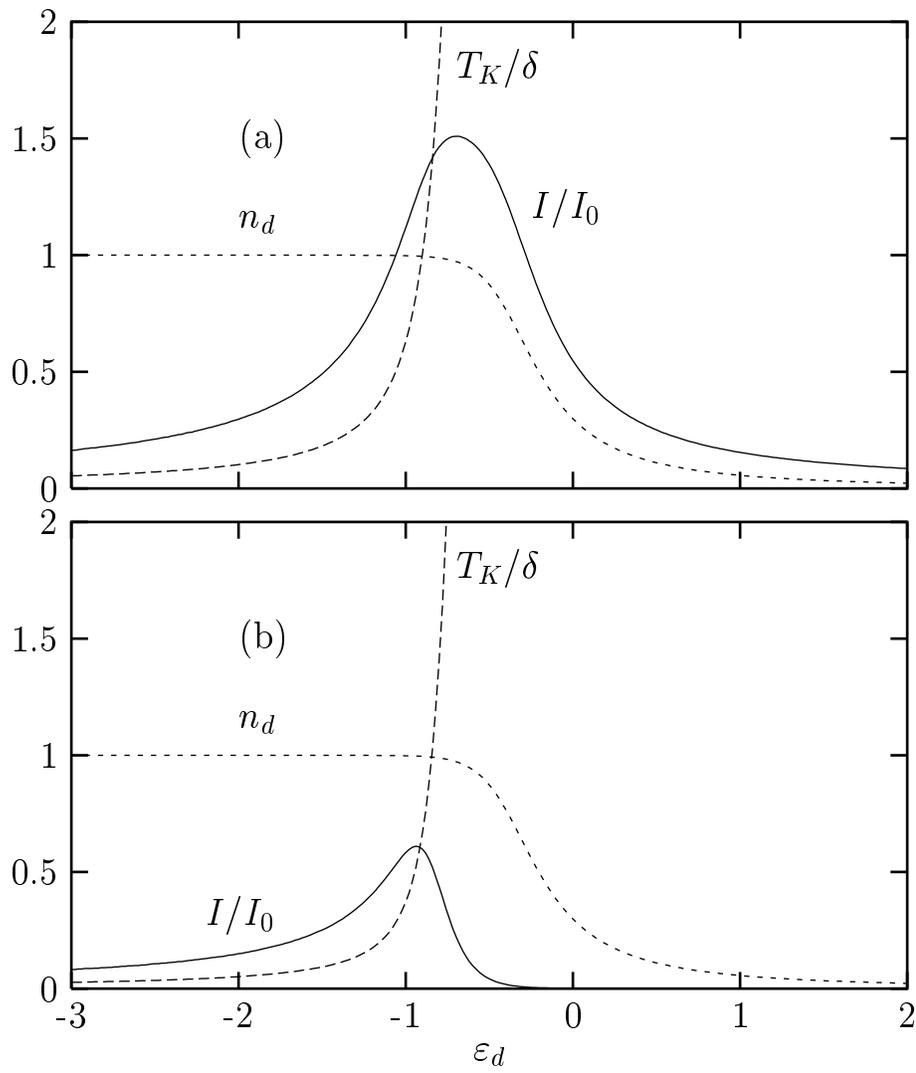
\end{document}